\documentclass[12pt,a4paper]{article}

\usepackage{graphicx}       
\usepackage{amsmath,amssymb}
\usepackage{geometry}       
\usepackage{cite}           
\usepackage{setspace}       

\title{\textbf{Role of Domain Walls in the Early Universe in the Context of Mode Matching}}
\author{
  K.K. Venkataratnam\\
  \small Department of Physics \\
  \small Malaviya National Institute of Technology Jaipur\\
  \small J.L.N. Marg, Jaipur, Rajasthan, India\\
  \small \texttt{kvkamma.phy@mnit.ac.in}
}

\begin{document}
\maketitle
\begin{abstract}
We study the role of domain walls and their relics in the very early Universe within the framework of the mode-matching technique. Domain walls formed during the spontaneous breaking of discrete symmetries are modelled as a short lived contribution to the background energy density, leading to a controlled deviation from standard slow-roll inflation. Solving the modified Friedmann equation, we obtain a smooth, time-dependent Hubble parameter that asymptotically approaches the standard inflationary value. We analyze the evolution of scalar perturbations using the gauge-invariant Mukhanov variable and perform mode matching across the transition between the domain wall affected phase and standard inflation. We find that only modes exiting the horizon near the transition experience a modified evolution, while modes far from the transition remain unaffected. As a result, the primordial scalar power spectrum exhibits localized, scale-dependent deviations from scale invariance, while preserving the overall success of inflation. 
These results demonstrate that unstable domain-wall relics can leave subtle imprints on primordial fluctuations, potentially providing a probe of early-Universe phase transitions.
\end{abstract}

\section{Introduction}

The inflationary paradigm provides a compelling explanation for the origin of the observed large-scale structure of the Universe \cite{Guth,Sato,Linde}. During an early phase of quasi-exponential expansion, quantum fluctuations of scalar fields were stretched beyond the Hubble radius, where they subsequently became classical and acted as seeds for cosmic inhomogeneities. One of the most robust predictions of inflation is a nearly scale-invariant primordial power spectrum, in excellent agreement with observations of the cosmic microwave background (CMB) \cite{Starobinsky,MukhanovChibisov,DurrerBook,MukhanovReview}.

The computation of primordial correlation functions in an expanding spacetime, however, involves conceptual and technical subtleties associated with quantum field theory in curved backgrounds \cite{BirrellDavies}. In a time-dependent Friedmann--Lema\^{i}tre--Robertson--Walker (FLRW) geometry, ultraviolet divergences arise in addition to infrared singularities, rendering the definition of physical observables nontrivial. These issues are particularly relevant for the power spectrum, which is directly related to the coincidence limit of the two-point function of the gauge-invariant Mukhanov variable $Q_k$ \cite{Mukhan,Mukhanov1988,MukhanovReview}. Ultraviolet divergences are commonly addressed through adiabatic subtraction in momentum space \cite{Parker1966,ParkerFulling}, which is equivalent to the DeWitt--Schwinger point-splitting method in coordinate space \cite{BunchParker,DeWitt}.

The physical relevance of adiabatic subtraction for observable quantities has been the subject of considerable debate. Parker and collaborators argued that adiabatic counterterms may leave residual imprints on the renormalized power spectrum near horizon exit. In contrast, Durrer, Marozzi, and Rinaldi demonstrated that such imprints are unphysical, emphasizing that the adiabatic expansion breaks down close to horizon crossing and that ultraviolet regularization should not affect observable spectra \cite{DurrerRinaldi}. Nevertheless, it has been noted that infrared effects associated with mode matching across different cosmological phases can potentially leave observable traces, provided the matching occurs sufficiently close to horizon exit \cite{JanssenProkopec,GlenzParker}.

The mode-matching technique has therefore emerged as a useful framework for probing deviations from standard slow-roll inflation. It has been applied to transitions between radiation-dominated and inflationary epochs \cite{JanssenProkopec}, to scenarios involving modified dispersion relations or lower-dimensional effective gravity \cite{Rinaldi2011}, and to the generation of gravitational waves during successive cosmological phases \cite{DurrerBook}. These studies indicate that even short-lived departures from standard inflationary evolution may imprint characteristic features on primordial spectra.

Closely related modifications of the expansion history can also arise during the reheating epoch following inflation. Recent studies have shown that reheating dynamics can significantly affect the number of $e$-folds, the effective equation of state, and the resulting observational constraints on inflationary models. In particular, detailed reheating analyses have been carried out for quadratic chaotic inflation \cite{YadavChaotic}, mutated hilltop inflation \cite{YadavHilltop}, and Mexican-hat–type potentials \cite{YadavMexicanHat}, highlighting the importance of post-inflationary physics in shaping primordial observables.Quantum fluctuations,density fluctuations and particle production in oscillatory and thermal inflaton back ground have also been investigated in related contexts\cite{Dhayal2020, Rathore2020, Dhwani2024}.

In this broader context, topological defects formed during early-Universe phase transitions provide a well-motivated source of localized deviations from standard cosmological evolution. Domain walls arise naturally from the spontaneous breaking of discrete symmetries in many particle-physics models \cite{Kibble,HindmarshKibble}. Although stable domain walls are cosmologically unacceptable due to their slow redshifting, several mechanisms can render them unstable or short-lived, allowing transient domain-wall networks to exist during pre-inflationary or early inflationary stages without conflicting with observations.

In this work, we investigate the role of transient domain walls and their relics in the very early Universe within the framework of the mode-matching technique. Focusing on the evolution of the gauge-invariant Mukhanov variable $Q_k$, we derive analytic solutions for both homogeneous and weakly inhomogeneous cosmological backgrounds. The effect of domain walls is modeled through a small perturbation in the Hubble parameter, $\Delta H$, capturing their leading-order influence on the background expansion. We analyze the resulting modifications to the primordial power spectrum and show that even short-lived domain-wall relics can induce mild, scale-dependent deviations from scale invariance, providing a potential probe of early-Universe phase transitions.

\section{Domain Wall Dynamics in an Expanding Universe}

\subsection{Formation of domain walls from discrete symmetry breaking}

Spontaneous symmetry breaking is a central concept in modern particle physics and cosmology \cite{Kibble,KolbTurner}. At sufficiently high temperatures in the early Universe, symmetries that are broken at late times are expected to be restored. As the Universe cools and undergoes phase transitions, discrete symmetries may break spontaneously, leading to the formation of topological defects such as domain walls, cosmic strings, and monopoles \cite{kibble1, VilenkinShellard}. Domain walls arise when the vacuum manifold is disconnected, and they correspond to two-dimensional topological defects separating regions of distinct but degenerate vacua.

At the time of a phase transition, the scalar field responsible for symmetry breaking acquires a nonzero vacuum expectation value (VEV) locally within causally disconnected regions. If the correlation length $\xi$ of the field is finite, regions separated by distances larger than $\xi$ choose different vacuum states independently. As a result, stable domain walls inevitably form at the boundaries between regions settling into different minima of the potential\cite{Kibble}. Once these regions enter the causal horizon, a network of domain walls with characteristic curvature scale of order the horizon size is expected to emerge.

\subsection{Field-theoretic realization and wall properties}

A simple and well-studied field-theoretic model that gives rise to domain walls is provided by a real scalar field $\phi$ with a discrete $Z_2$ symmetry, described by the Lagrangian density
\begin{equation}
\mathcal{L} = \frac{1}{2}\partial_\mu \phi\, \partial^\mu \phi
- \frac{\lambda}{4}(\phi^2 - \sigma^2)^2 ,
\end{equation}
where $\lambda$ and $\sigma$ are positive constants. The potential possesses two degenerate minima at $\phi = \pm \sigma$, and the spontaneous breaking of the $Z_2$ symmetry leads to the formation of domain walls interpolating between these vacua.

For a static planar configuration depending only on a single spatial coordinate $z$, the equation of motion admits the well-known kink solution
\begin{equation}
\phi_w(z) = \sigma \tanh\!\left(\frac{z}{\Delta}\right),
\end{equation}
where the characteristic thickness of the wall is given by
\begin{equation}
\Delta = \left(\sqrt{\frac{\lambda}{2}}\,\sigma\right)^{-1}.
\end{equation}
The wall thickness is determined by the balance between gradient energy, which favors thinner walls, and potential energy, which favors broader configurations.

The energy--momentum tensor associated with the scalar field takes the standard form
\begin{equation}
T_{\mu\nu} = \partial_\mu \phi\,\partial_\nu \phi - g_{\mu\nu}\mathcal{L}.
\end{equation}
Substituting the wall solution yields a localized energy density
\begin{equation}
T_{00} = \frac{\lambda}{2}\sigma^4 \,
\text{sech}^4\!\left(\frac{z}{\Delta}\right),
\end{equation}
from which the surface energy density (or tension) of the wall is obtained as
\begin{equation}
\sigma_{\rm wall} = \int dz\, T_{00}
= \frac{2\sqrt{2}}{3}\lambda^{1/2}\sigma^3 .
\end{equation}
The equality of surface energy density and tension reflects the inherently relativistic nature of domain walls.

\subsection{Cosmological evolution and equation of state}

Following their formation, domain walls evolve as a network embedded in an expanding FLRW background. Averaged over large scales, a domain-wall network can be characterized by an effective equation of state. For walls moving with rms velocity $v$, the pressure $P_w$ and energy density $\rho_w$ satisfy \cite{Vilenkin}
\begin{equation}
P_w = \left(v^2 - \frac{2}{3}\right)\rho_w .
\end{equation}
In the non-relativistic limit $v \rightarrow 0$, this corresponds to an equation-of-state parameter $w = -2/3$, implying that the energy density of domain walls scales as $\rho_w \propto a^{-1}$.

This extremely slow redshifting causes stable domain walls to quickly dominate the energy density of the Universe, leading to severe conflicts with standard cosmology. Consequently, viable cosmological scenarios require domain walls to be unstable or short-lived.The evolution and decay of unstable domain-wall networks have been studied analytically through numerical simulations in various cosmological settings\cite{Matsuda,GWdomainwalls1,GWdomainwalls2}. Such transient domain walls may arise if the degeneracy between vacua is lifted by a small explicit symmetry-breaking term, allowing the walls to decay before dominating the expansion. Despite their instability, these domain walls can still influence the background dynamics over a finite period, motivating their study in the context of early-Universe cosmology.

\section{Domain Walls and Inflationary Background Evolution}

The presence of domain walls in the early Universe modifies the background expansion through their contribution to the total energy density. In a spatially flat Friedmann--Lema\^{i}tre--Robertson--Walker spacetime, the evolution of the scale factor $a(t)$ is governed by the Friedmann equation,
\begin{equation}
H^2 = \frac{8\pi G}{3}\rho,
\end{equation}
where $H=\dot a/a$ is the Hubble parameter and $\rho$ denotes the total energy density.

In the scenario considered here, the energy density receives contributions from the inflaton vacuum energy, $\rho_\Lambda$, and from a transient network of domain walls, $\rho_{DW}$. Owing to their effective equation of state $w=-2/3$, the domain-wall energy density scales as $\rho_{DW}\propto a^{-1}$, while the inflaton vacuum energy remains approximately constant. The Friedmann equation may therefore be written as
\begin{equation}
H^2 = \frac{8\pi G}{3}
\left(
\rho_\Lambda + \frac{\rho_{DW}\,a_i}{a}
\right),
\end{equation}
where $a_i$ denotes the scale factor at the epoch of domain-wall formation.

The above equation admits an analytic solution for the scale factor, which can be expressed as
\begin{equation}
a(t) = \frac{c_3}{c_2}
\cosh^2\!\left[\frac{\sqrt{c_2}}{2}(t-t_i)\right],
\end{equation}
with
\begin{equation}
c_2 = \frac{8\pi G}{3}\rho_\Lambda \equiv H_0^2,
\qquad
c_3 = \frac{8\pi G}{3}\rho_{DW}a_i .
\end{equation}
At late times, when the contribution from domain walls becomes negligible, the scale factor asymptotically approaches the standard inflationary solution $a(t)\propto e^{H_0 t}$.

\subsection{Motivation for mode matching}

The transition from a domain-wall--affected phase to a pure slow-roll inflationary phase occurs over a finite time interval during which the Hubble parameter varies smoothly. For modes whose wavelengths are comparable to the Hubble radius during this transition, the evolution of quantum fluctuations may deviate from the standard slow-roll behavior.

To capture these effects in a controlled manner, we employ the mode-matching technique. The cosmological evolution is divided into two regimes: an early phase in which the domain-wall contribution modifies the background dynamics, and a later phase described by quasi–de Sitter inflation. At a matching time $t=t_1$, chosen such that the domain-wall energy density becomes subdominant, the scale factor and Hubble parameter are continuous, ensuring a consistent evolution of perturbations across the transition.

This approach allows us to isolate the impact of transient domain-wall relics on the evolution of scalar perturbations while preserving the successful predictions of standard inflation at late times.

 \section{Scalar Perturbations and Mode Matching in the Presence of Domain Walls}

We now study the evolution of scalar perturbations on the background modified by transient domain-wall contributions.The gauge-invariant treatment of cosmological perturbations has been extensively developed in the literature\cite{MukhanovReview1,MukhanovReview}. The relevant gauge-invariant quantity describing scalar perturbations is the Mukhanov variable $Q_k$, which combines inflaton and metric fluctuations into a single canonical degree of freedom \cite{Mukhanov1988,MukhanovReview}. It is defined as
\begin{equation}
Q_k \equiv a\left(\delta\phi_k + \frac{\dot{\phi}}{H}\psi_k\right),
\end{equation}
where $\delta\phi_k$ denotes the inflaton fluctuation and $\psi_k$ represents the scalar metric perturbation in longitudinal gauge. The dynamics of $Q_k$ fully encodes the generation and evolution of the primordial scalar power spectrum.

In a general time-dependent background, the Mukhanov variable satisfies the equation of motion
\begin{equation}
\ddot{Q}_k + 3H\dot{Q}_k +
\left[
\frac{k^2}{a^2}
+ H^2 (3\eta - 6\epsilon + 2\epsilon^2)
- \frac{\ddot{H}}{H}
\right] Q_k = 0 ,
\end{equation}
where $\epsilon=-\dot H/H^2$ and $\eta$ are the usual slow-roll parameters. The presence of domain walls modifies the Hubble parameter $H(t)$, thereby inducing corrections to the effective mass term of the perturbation equation.

It is often convenient to express the dynamics in conformal time $\tau$, in which case the perturbation equation takes the canonical Mukhanov--Sasaki form
\begin{equation}
Q_k'' + \left(k^2 - \frac{z''}{z}\right) Q_k = 0 ,
\qquad
z \equiv \frac{a\dot{\phi}}{H},
\end{equation}
where primes denote derivatives with respect to conformal time. Deviations from slow-roll inflation induced by the transient domain-wall phase modify the quantity $z''/z$, leading to scale-dependent corrections in the evolution of $Q_k$.

To capture the effect of the domain-wall epoch, we implement a mode-matching procedure at the transition time $t_1$ between the domain-wall–dominated phase and the standard inflationary phase. Continuity of the perturbations requires the matching conditions
\begin{equation}
Q_k^{(DW)}(t_1) = Q_k^{(\mathrm{inf})}(t_1),
\qquad
\dot{Q}_k^{(DW)}(t_1) = \dot{Q}_k^{(\mathrm{inf})}(t_1),
\end{equation}
which determine how the pre-inflationary domain-wall dynamics imprints itself on the inflationary perturbation modes.
Using the background solution derived in the previous section,
\begin{equation}
H(t) = H_0 \tanh\!\left[\frac{H_0}{2}(t-t_i)\right].
\end{equation}
Evolution of Hubble parameter H versus time(t) is as shown in Figure 1.

\begin{figure}
    \centering
    \includegraphics[width=0.5\linewidth]{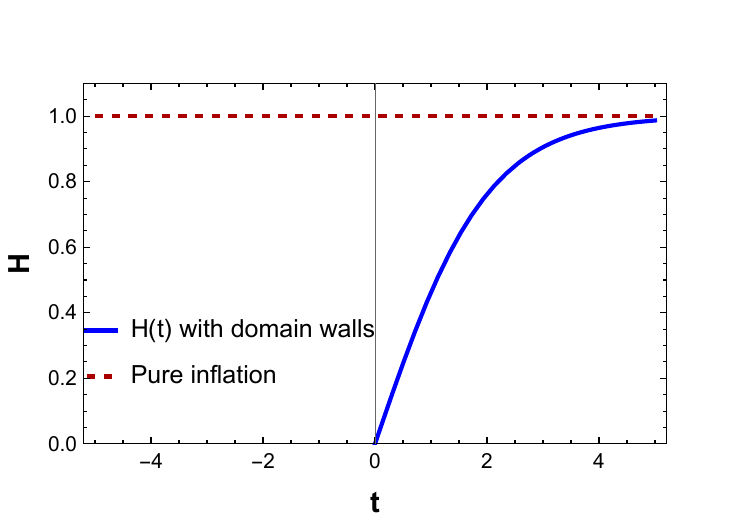}
    \caption{Evolution of the Hubble parameter in the presence of a transient domain-wall contribution. The deviation from constant $H_0$ is localized around the matching time, after which standard slow-roll inflation is recovered.}
    \label{fig1}
\end{figure}

Figure 1 illustrates the evolution of the Hubble parameter in the presence of a transient domain-wall contribution. The deviation from constant slow-roll expansion is smooth and short-lived, reflecting the rapid redshifting of the domain-wall energy density. Standard inflationary evolution is fully recovered at late times.

The first slow-roll parameter at the matching time $t=t_1$ is given by
\begin{equation}
\epsilon(t_1) =
-\frac{\dot H}{H^2}\Bigg|_{t=t_1}
=
\frac{1}{2}
\coth^2\!\left[\frac{H_0}{2}(t_1-t_i)\right].
\end{equation}
In standard slow-roll inflation, $\epsilon_{\rm inflation}\ll1$ is approximately constant, and the deviation induced by the transient domain-wall phase can therefore be written as
\begin{equation}
\Delta\epsilon(t_1) = \epsilon(t_1)-\epsilon_{\rm inflation}.
\end{equation}

Similarly, the second slow-roll parameter,
\begin{equation}
\eta \equiv \frac{\dot\epsilon}{\epsilon H},
\end{equation}
evaluated at the matching time yields
\begin{equation}
\eta(t_1)=1-\frac{1}{2}
\coth^2\!\left[\frac{H_0}{2}(t_1-t_i)\right],
\end{equation}
leading to a corresponding deviation $\Delta\eta(t_1)=\eta(t_1)-\eta_{\rm inflation}$. These departures from slow-roll behavior modify the effective mass term in the Mukhanov equation and play a crucial role in determining the resulting power spectrum.

\begin{figure}
    \centering
    \includegraphics[width=0.5\linewidth]{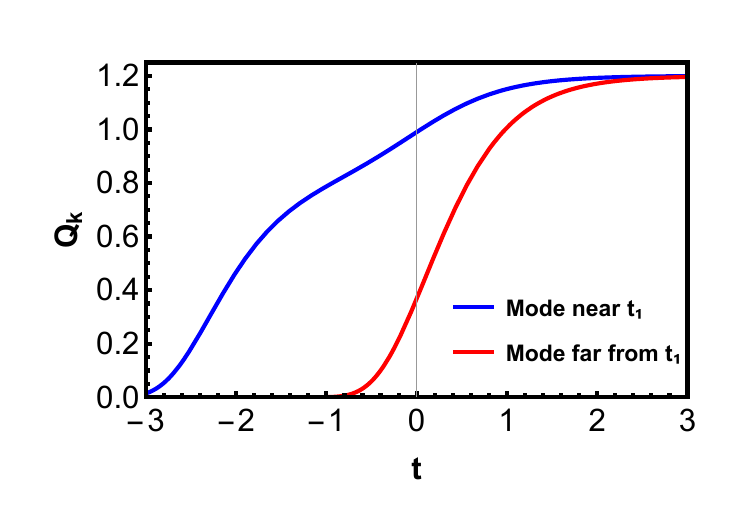}
    \caption{Time evolution of the Mukhanov variable $|Q_k|$ for representative modes. Modes exiting the horizon near the matching time exhibit deviations due to the domain-wall phase, while modes far from the transition remain unaffected.}
    \label{fig2}
\end{figure}

Mukhanov variable $Q_k$ versus time(t) is as shown in Figure 2.
Figure 2 shows the evolution of the Mukhanov variable $Q_k$ for representative modes. While modes exiting the horizon far from the transition behaves as in standard inflation, modes exiting near the transition experience a temporary distortion. This highlights the loacalized impact of the domain-wall phase on scalar perturbations.

\section{Primordial Power Spectrum and Domain-Wall Signatures}

The observable imprint of transient domain walls is encoded in the primordial scalar power spectrum, which is determined by the late-time behavior of the Mukhanov variable $Q_k$. For single-field inflation, the dimensionless power spectrum of curvature perturbations may be written as \cite{MukhanovReview,LiddleLyth}
\begin{equation}
P_\epsilon(k) =
\frac{k^3}{2\pi^2}
\left(\frac{H}{\dot{\phi}}\right)^2
|Q_k|^2
=
\frac{k^3}{4\pi^2 M_p^2 \epsilon}
|Q_k|^2 ,
\end{equation}
where $\epsilon$ is the slow-roll parameter and $M_p$ is the reduced Planck mass.

The presence of a transient domain-wall phase modifies both the background slow-roll parameters and the mode functions. To leading order, the fractional change in the power spectrum may be expressed as
\begin{equation}
\frac{\Delta P_\epsilon(k)}{P_\epsilon(k)} =
\frac{2\,\Delta |Q_k|}{|Q_k|}
-
\frac{\Delta\epsilon}{\epsilon} ,
\end{equation}
where the first term arises from corrections to the Mukhanov mode amplitude due to mode matching, while the second term reflects deviations from slow-roll evolution induced by the domain-wall contribution to the Hubble parameter.

Since the domain-wall energy density scales as $\rho_{DW}\propto a^{-1}$, its influence is confined to early times and rapidly becomes negligible during inflation. As a result, only modes whose wavelengths are comparable to the Hubble radius near the matching time $t_1$ are significantly affected. Modes exiting the horizon well before or well after this epoch retain the nearly scale-invariant spectrum predicted by standard slow-roll inflation.
The fractional change in power spectrum $\frac{\Delta P}{P} $ versus Mode $k$ is as shown in Figure 3.

\begin{figure}
    \centering
    \includegraphics[width=0.5\linewidth]{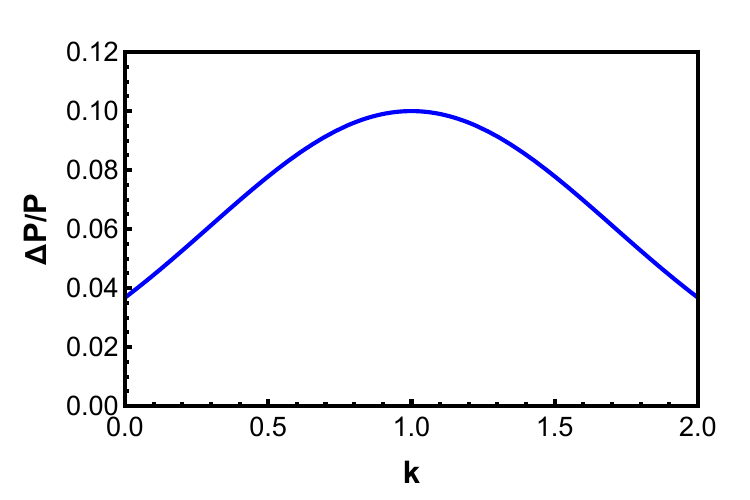}
    \caption{Fractional modification of the primordial scalar power spectrum induced by a transient domain-wall phase. Deviations are localized around modes exiting the horizon near the matching time, while scale invariance is recovered at larger and smaller scales.}
    \label{fig3}
\end{figure}

The modification is localized around a characteristic scale corresponding to horizon exit near the transition time, while scale invariance is preserved at both larger and smaller scales. Such features are compatible with current observational constraints and may be relevant for large-scale anomalies.

The resulting power spectrum therefore exhibits localized, scale-dependent deviations from scale invariance, rather than a global distortion. Such features are qualitatively similar to those produced by brief interruptions of slow-roll inflation or by pre-inflationary relics. Importantly, because the domain walls are assumed to be unstable and short-lived, the overall success of inflation is preserved, while allowing for potentially observable imprints at large scales.
 
\section{Discussion and Conclusions}
In this work, we have examined the cosmological implications of a transient domain-wall phase in the very early Universe and its impact on inflationary dynamics. Modeling the domain-wall network as a short-lived contribution to the background energy density, we derived the modified expansion history and demonstrated that the resulting deviation from slow-roll inflation is smooth and rapidly diluted. Using a mode-matching approach to solve the Mukhanov–Sasaki equation, we quantified how this temporary modification affects scalar perturbations. We find that only modes exiting the horizon near the transition epoch experience a measurable distortion, while modes well before or after the event evolve as in standard slow-roll inflation.

The imprints of this transient phase appear as localized, scale-dependent features in the primordial scalar power spectrum, with amplitudes consistent with current observational bounds. The effects are confined to a narrow band of scales, preserving near scale invariance elsewhere and maintaining the overall robustness of inflationary predictions. Our results highlight how short-lived relics of symmetry breaking can leave subtle yet potentially observable signatures in primordial perturbations. This framework offers a systematic pathway for connecting early-Universe phase transitions with precision cosmological data and motivates further exploration of tensor modes, non-Gaussianities, and concrete particle-physics realizations of unstable domain-wall networks.

\section{Acknowledgments}
The author gratefully acknowledges Prof. Urjit A. Yajnik for his valuable guidance and constructive discussions during the course of this work. The author also thanks Sudhava Yadav for her assistance in preparing the manuscript and useful discussions.

 \end{document}